\documentclass{appolb}
\usepackage{epsfig}

\def\bge{\begin{equation}}
\def\ene{\end{equation}}
\def\bg{\begin{eqnarray}}
\def\en{\end{eqnarray}}

\def\bge{\begin{equation}}
\def\ene{\end{equation}}
\def\bg{\begin{eqnarray}}
\def\en{\end{eqnarray}}

\begin{document}
\title{
$\eta'$ interactions with nucleons and nuclei
}
\author{Steven D. Bass 
\address{Stefan Meyer Institute for Subatomic Physics, \\
Austrian Academy of Sciences,
Boltzmanngasse 3, 1090 Vienna, Austria}
\and
Pawel Moskal
\address{
Institute of Physics, Jagiellonian University, \\
ul. Lojasiewicza 11, PL 30-348 Cracow, Poland}
}
\maketitle
\begin{abstract}
\noindent
We summarise recent progress in theory and experiment 
towards understanding $\eta'$ meson interactions with 
nucleons and nuclei. 
Highlights include the production mechanism of $\eta'$ 
mesons in proton-proton collisions close to threshold,
the $\eta'$ effective mass shift in nuclei and 
the determination of the $\eta'$-nucleon scattering 
length in free space.
\end{abstract}

\section{Introduction}

The last 20 years have witnessed a dedicated programme of 
experimental and theoretical studies of low-energy $\eta'$ interactions with nucleons, nuclei and other mesons.
The $\eta'$ meson is special in QCD because of its strong 
affinity to glue. 
While pions and kaons are would-be Goldstone bosons associated 
with chiral symmetry, the isosinglet $\eta$ and $\eta'$ mesons 
are too massive by about 300-400 MeV for them to be pure 
Goldstone states. 
They receive extra mass from non-perturbative gluon dynamics associated with the QCD axial anomaly.
Taking the $\eta$--$\eta'$ mixing angle between $-15^\circ$ and $-20^\circ$, the $\eta'$ is predominantly a 
flavour-singlet state with strong coupling to gluonic intermediate states meaning that its interactions with 
other hadrons are, in general, characterised by OZI violation,
for recent reviews see \cite{bt2013,zakopane,shore}.
The experimental programme has focussed on near threshold
$\eta'$ production in proton-nucleon collisions using the 
COSY-11 facility at FZ-J\"ulich \cite{pawelcosy11},
$\eta'$ photoproduction experiments at ELSA in Bonn 
\cite{elsaproton} and Jefferson Laboratory \cite{jlabproton}, 
studies of the $\eta'$ in medium 
and the search for $\eta'$ bound states in nuclei at 
ELSA, GSI and LEPS2 \cite{metagexa} and 
production of hadronic states with exotic 
quantum numbers at COMPASS at CERN \cite{compassexotic}.

Highlights from COSY-11 \cite{pawelcosy11,pawelrev}
include studies of the $\eta'$ and $\eta$ 
production mechanisms in proton-nucleon collisions close
to threshold through measurements of the total
and differential cross-sections and varying the isospin of 
the second nucleon.
The $\eta'$ is observed to be produced primarily in 
s-wave up to excess energy of at least ${\cal E} =11$ MeV.
A first quantitative value of the 
$\eta'$-nucleon scattering length has been obtained 
\cite{eryk2014}
as well
as the most accurate measurement of the $\eta'$ total width 
in free space \cite{erykwidth}.
Photoproduction measurements \cite{elsaproton,jlabproton}
from proton and deuteron targets have recently been 
extended by the CBELSA/TAPS 
collaboration in Bonn to carbon and niobium to make a 
first (indirect) measurement of the $\eta'$-nucleus
optical potential \cite{nanova,metagexa}. 
One finds an $\eta'$ effective mass shift in nuclei of
about -37 MeV at nuclear matter density, in excellent agreement with the prediction of the Quark Meson Coupling
model \cite{etaA}, QMC, through coupling of 
the light quarks in the 
meson to the $\sigma$ mean field inside the nucleus.
The $\eta'$ experiences an effective mass shift in 
nuclei which is catalysed by its gluonic component
\cite{bt2013,exa14}.
Although the $\eta'$-nucleon interaction 
in free space is much weaker \cite{eryk2014} 
than the $\eta$-nucleon interaction \cite{wycech}, 
the small width of the $\eta'$ in medium \cite{elsa}
means that the $\eta'$ may be a good candidate 
for possible bound state searches,
{\it e.g.} in experiments at ELSA, GSI and LEPS2 \cite{metagexa}.
Searches for $\eta$ mesic nuclei are ongoing with data 
from WASA-at-COSY \cite{wasaeta}.
The odd $L$ exotic partial waves $L^{-+}$ are 
strongly enhanced in $\eta' \pi$ relative to $\eta \pi$ 
exclusive production in collisions of 
191 GeV negatively charged pions from hydrogen at COMPASS, 
consistent with expectations based on the axial U(1) 
extended chiral Lagrangian \cite{bassmarco}.

\section{QCD symmetries and the $\eta$ and $\eta'$}

Spontaneous chiral symmetry breaking in QCD induces an octet 
of Goldstone bosons associated with SU(3) and also 
(before extra gluonic effects in the singlet channel)
a flavour-singlet Goldstone boson.
The mass squared of these Goldstone bosons 
is proportional to the current mass of their valence quarks.
While the pion and kaon fit well in this picture,
to understand the isosinglet $\eta$ and $\eta'$ masses
one needs 
300-400 MeV extra mass in the flavour-singlet channel 
which is associated with non-perturbative topological 
gluon configurations \cite{shore,crewther}
related perhaps to confinement \cite{ks} 
or instantons \cite{thooft}.
The gluonic mass contribution ${\tilde m}^2_{\eta_0}$
satisfies the Witten-Veneziano mass formula 
\cite{witten,venez}
\begin{equation}
m_{\eta}^2 + m_{\eta'}^2 = 2 m_K^2 + {\tilde m}_{\eta_0}^2 
\end{equation}
and has a rigorous interpretation in terms of 
the QCD Yang-Mills topological susceptibility. 
SU(3) breaking generates mixing between the octet and singlet
states which, together with the gluonic mass term, yields the massive $\eta$ and $\eta'$ bosons.

Phenomenological studies of various decay processes give a value for the $\eta$-$\eta'$ mixing angle between $-15^\circ$ and $-20^\circ$ \cite{gilman}.
The $\eta'$ has a large flavour-singlet component with strong affinity to couple to gluonic degrees of freedom, 
{\it e.g.} in OZI violating interactions.
For the $\eta$ meson the singlet component is also important,
particularly in understanding the $\eta$ in nuclei and 
potentially accounting for 50\% of the $\eta$-nucleon scattering length in medium \cite{etaA}.
In the OZI limit of no gluonic mass term the $\eta$ would 
be approximately an isosinglet light-quark state
(${1 \over \sqrt{2}} | {\bar u} u + {\bar d} d \rangle$)
with mass $m_{\eta} \sim m_{\pi}$
degenerate with the pion and the $\eta'$ would be a 
strange-quark state $| {\bar s} s \rangle$
with mass $m_{\eta'} \sim \sqrt{2 m_{\rm K}^2 - m_{\pi}^2}$,
mirroring the isoscalar vector $\omega$ and $\phi$ mesons.

The gluonic mass term is related to the QCD axial anomaly 
in the divergence of the flavour-singlet axial-vector current.
While the non-singlet axial-vector currents are partially conserved (they have just mass terms in the divergence), the singlet current
$
J_{\mu 5} = \bar{u}\gamma_\mu\gamma_5u
+ \bar{d}\gamma_\mu\gamma_5d + \bar{s}\gamma_\mu\gamma_5s 
$
satisfies the anomalous divergence equation 
\begin{equation}
\partial^\mu J_{\mu5} = 6 Q
+ \sum_{k=1}^{3} 2 i m_k \bar{q}_k \gamma_5 q_k 
\end{equation}
where 
$
Q = \partial^{\mu} K_{\mu}
= {\alpha_s \over 8 \pi} G_{\mu \nu} {\tilde G}^{\mu \nu}
$
is the topological charge density.
The integral over space $\int \ d^4 z \ Q = n$ measures the 
gluonic winding number \cite{crewther} 
which is an integer for (anti-)instantons and which vanishes 
in perturbative QCD.

The anomalous glue that generates the large $\eta$ and $\eta'$ 
masses also drives OZI violating $\eta$ and $\eta'$ production 
and decay processes \cite{bassmarco,gilman,veccb,hf}
and enters in the $\eta'$-nucleon interaction \cite{bass99}.
In high energy processes $B$ and charm-quark meson decays involving an $\eta'$ in the final state are driven in part 
by strong coupling to gluonic intermediate states
\cite{gilman,hf}.
In low energy QCD the $\eta'$ experiences an effective mass
shift in nuclei that, 
within the QMC model,
is catalysed by its gluonic component \cite{bt2013}.
The $\eta'$-nucleon coupling constant is, 
in principle, sensitive to OZI violation \cite{bass99}. 
The QCD axial anomaly also 
plays an important role in the interpretation of the 
flavour-singlet Goldberger-Treiman relation \cite{tgv}
and the nucleon's flavour-singlet axial-charge 
(or ``quark spin content'')
measured in polarised deep inelastic scattering 
associated with the proton spin puzzle \cite{SBrmp1,SBrmp2}.
We refer to Ref.~\cite{sbcracow} for a discussion of gluonic components in the $\eta'$ wavefunction and mixing with pseudoscalar glueball states.

The axial U(1) extended chiral Lagrangian \cite{vecca}
incorporates the chiral and axial U(1) symmetries and 
allows us to study low-energy QCD processes involving 
the $\eta'$.
The gluonic mass term ${\tilde m}_{\eta_0}^{2}$ 
is introduced via a flavour-singlet potential involving the topological charge density $Q$ which is constructed so that the Lagrangian also reproduces the axial anomaly \cite{vecca}.
Potential terms involving $Q$ generally describe OZI violation,
{\it e.g.}  
the term 
$Q^2 \partial_{\mu} \pi_a \partial^{\mu} \pi_a$
with $\pi_a$ the pseudoscalar Goldstone fields 
drives the decay $\eta' \rightarrow \eta \pi \pi$
\cite{veccb}
and plays an important role in dynamical generation of a
light mass exotic with quantum numbers $1^{-+}$ 
\cite{bassmarco}, see below.

\section{$\eta'$ production experiments}

$\eta'$ production has been measured in proton-proton
collisions close to threshold 
(excess energy ${\cal E}$ between 0.76 and $\sim 50$ MeV) 
by the COSY-11 collaboration at FZ-J\"ulich 
[33 - 37]
and at ${\cal E} = 3.7$ MeV and 8.3 MeV 
by SPESIII \cite{hibou} 
and 
144 MeV by the DISTO Collaboration at SATURNE \cite{disto}.

For the $\eta'$, production is s-wave dominated for ${\cal E}$
up to at least 11 MeV.
The proton-proton and $\eta'$-proton invariant mass distributions determined for the $pp \rightarrow pp \eta'$
reaction at excess energy ${\cal E}=16.4$ MeV show an enhancement
which might indicate a non-negligible p-wave contribution
from the proton-proton subsystem \cite{pklaja}.
Fitting the low ${\cal E}$ data to models of 
the $\eta'$ final state interaction allowed 
COSY-11 to extract a first measurement of the 
$\eta'$-proton scattering length \cite{eryk2014}, see Eq.(4) below.

Comparison of $\pi^0$, $\eta$ and $\eta'$ 
production in proton-nucleon collisions close to threshold
was performed at COSY-11.
For near-threshold meson production, the production 
cross-section is reduced by initial state interaction
between the incident nucleons and enhanced by final state
interactions between the outgoing hadrons. 
For comparing production dynamics a natural variable is
the volume of available phase space 
which is approximately independent of the meson mass.
Making this comparison for the neutral pseudoscalar mesons, 
it was found that production of the $\eta$ meson is about 
six times enhanced compared to the $\pi^0$ which is six 
times further enhanced compared to the $\eta'$ \cite{paweletapc}.
One may conclude that the production of the $\eta'$ and
$\pi^0$ close to threshold is non-resonant in contrast 
to $\eta$ production which proceeds through strong coupling 
to $S_{11} (1535)$ \cite{faeldt}. 
However, it should be noted that as advocated in 
Ref.~\cite{kanzo},
$\eta'$ meson production may also be explained by the 
relatively weak coupling to a rather not well established 
set of s-wave and p-wave resonances.
Based on the comparison of excitation functions for the 
$pp \rightarrow pp \eta$ and $pp \rightarrow pp \eta'$ reactions close to threshold
it was concluded that the $\eta$-proton interaction 
is much stronger than for $\eta'$-proton \cite{paweletapc}.
In higher energy experiments with proton-proton collisions 
at 450 GeV the $\eta$ and $\eta'$ seem to have a similar production mechanism which differs from that of the $\pi^0$ \cite{heetap}.

Measurements of the isospin dependence of $\eta$ meson production in proton-nucleon collisions 
revealed that the total cross-section 
for the quasi-free $pn \rightarrow pn \eta$ reaction
exceeds the corresponding cross section 
for $pp \rightarrow pp \eta$ 
by a factor of about three at threshold and by factor of 
six at higher excess energies between about 25 and 100 MeV
\cite{pawela,uppsala}. 
Combining information about the strong isospin dependence and 
the isotropic angular distributions of the $\eta$ meson emission angle in the centre-of-mass frame, it was established that the $\eta$ meson is predominantly created via excitation of one of the nucleons to the $S_{11}$(1535) resonance via a strong isovector exchange contribution. 
The angular dependence of the analysing power slightly
indicated that the process proceeds via exchange of the 
$\pi$ meson \cite{paweleta}.

Measurements of the isospin dependence of $\eta'$ 
production further suggest a different production
mechanism for this meson \cite{paweletapc,pawelb}. 
Using the quasi-free proton-neutron interaction \cite{qfree}
COSY-11 placed an upper bound on 
$\sigma (pn \rightarrow pn \eta')$ 
and the ratio
$R_{\eta'} = 
\sigma (pn \rightarrow pn \eta') 
/ \sigma(pp \rightarrow pp \eta')$ \cite{pawelb}.
For excess energy between 8-24 MeV $R_{\eta'}$
was observed to be consistently one standard 
deviation below the corresponding ratio for $\eta$
production \cite{pawela}. 
In the gendanken limit that $\eta'$ production proceeded
entirely through gluonic excitation in the intermediate
state this ratio would go to one. The data is consistent
with both a role for OZI violating $\eta'$ production 
\cite{bass99} and the meson exchange model \cite{kampfer}.
The data do not favour a dominant role for the $S_{11}$(1535)
in the $\eta'$ production mechanism, unlike for $\eta$
production.

As an extra bonus from these experiments, the total width
of the $\eta'$ was determined from its mass distribution
to be 
$\Gamma = 0.226 \pm 0.017 (stat.) \pm 0.014 (syst.)$ MeV
\cite{erykwidth},
an order of magnitude more accurate than previous determinations.

$\eta'$ (quasi-free) photoproduction from proton and 
deuteron targets was studied at ELSA \cite{elsaproton}
and JLab \cite{jlabproton}.
The production cross-section is isospin independent
for incident photon energies greater than 2 GeV, 
where t-channel exchanges are important.
At lower energies, particularly between 1.6 and 1.9 GeV
where the proton cross-section peaks, the proton and
quasi-free neutron cross-sections show different
behaviour, perhaps associated with resonances or
interference terms \cite{elsaproton}.

\section{The $\eta$ and $\eta'$ in nuclei}

Recent progress in theoretical and experimental studies of 
the $\eta-$ and $\eta'-$ (as well as pion and kaon)
nucleus systems promises to yield valuable new information about dynamical chiral and axial U(1) symmetry breaking 
in low energy QCD \cite{bt2013}.
With increasing nuclear density chiral symmetry is 
partially restored corresponding to a reduction in 
the values of the quark condensate and pion decay constant 
$f_{\pi}$ \cite{kienle,suzuki}.
This in turn leads to changes in the properties of hadrons 
in medium including the masses of the Goldstone bosons. 
There is presently vigorous experimental 
[7,12,16,17,51 - 58]
and theoretical 
[1,13,59 - 64]
activity aimed at understanding 
the $\eta$ and $\eta'$ in medium and 
to search for evidence of possible
$\eta$ and $\eta'$ bound states in nuclei.
How does the gluonic part of their mass change in nuclei?
Medium modifications need to be understood 
self-consistently within the interplay of confinement,
spontaneous chiral symmetry breaking and axial U(1) dynamics.

The $\eta$- and $\eta'$-nucleon interactions are believed 
to be attractive corresponding to a reduced effective mass 
in the nuclear medium and the possibility that 
these mesons might form strong-interaction bound-states in nuclei.
For the $\eta$ one finds a sharp rise in the cross section at
threshold for $\eta$ production in both
photoproduction from $^3$He \cite{mamieta}
and in proton-deuteron collisions \cite{pdeta}
which may hint at a reduced $\eta$ effective mass 
in the nuclear medium.
Measurement of the $\eta'$-nucleus optical potential by 
the CBELSA/TAPS collaboration suggests that the effective 
$\eta'$ mass drops by about 40 MeV at nuclear matter density
\cite{nanova}.
For the pion and kaon systems one finds a small pion 
mass shift of order a few MeV in nuclear matter \cite{kienle}
whereas kaons are observed to experience an effective 
mass drop for the $K^-$ to about 270 MeV at two times 
nuclear matter density in heavy-ion collisions 
\cite{gillitzer,kaons}.
The same heavy-ion experiments also suggest the effective 
mass of anti-protons is reduced by about 100-150 MeV 
below their mass in free space \cite{gillitzer}.
Experiments in heavy-ion collisions \cite{averbeck} and $\eta$ photoproduction from nuclei \cite{robig,yorita} suggest little modification of the $S_{11}(1535)$ excitation in-medium, 
though some evidence for the broadening of the $S_{11}$ 
in nuclei was reported in Ref.~\cite{yorita}.

Building on $\eta'$ photoproduction from proton targets,
meson mass shifts in medium can be investigated through 
studies of excitation functions in photoproduction 
experiments from nuclear targets
and through searches for possible meson bound states in nuclei. 
In photoproduction experiments
the production cross section is enhanced with the lower effective meson mass in the nuclear medium. 
When the meson leaves the nucleus it returns on-shell 
to its free mass with the energy budget conserved at
the expense of the kinetic energy so that excitation functions
and momentum distributions can provide essential clues to the
meson properties in medium \cite{metag}.
Using this physics a first (indirect) estimate of the $\eta'$ 
mass shift has recently been deduced 
by the CBELSA/TAPS Collaboration \cite{nanova}.
The $\eta'$-nucleus optical potential 
$V_{\rm opt} = V_{\rm real} + iW$
deduced from these photoproduction experiments is 
\begin{eqnarray}
V_{\rm real} (\rho_0)
= m^* - m 
&=& -37 \pm 10 (stat.) \pm 10 (syst.) \ {\rm MeV}
\nonumber \\ 
W(\rho_0) &=& -10 \pm 2.5 \ {\rm MeV}
\end{eqnarray}
at nuclear matter density $\rho_0$.
In this experiment the average momentum of the produced 
$\eta'$ was 1.1 GeV and the mass shift was measured in
production from a carbon target.
This optical potential corresponds to an effective
scattering length in medium with real part about 0.5 fm 
assuming we switch off the Ericson-Ericson rescattering 
denominator \cite{ericson}.

The COSY-11 collaboration have recently determined 
the $\eta'$-nucleon scattering length in free space
to be
\begin{eqnarray}
\nonumber
\mathrm{Re}(a_{\eta' p}) &=&  0~\pm~0.43~\mathrm{fm}
\\
\mathrm{Im}(a_{\eta' p}) &=& 0.37^{~+0.40}_{~-0.16}~\mathrm{fm}\end{eqnarray}
from studies of the final state interaction in $\eta'$ production in proton-proton collisions close to threshold 
\cite{eryk2014}.
Theoretical models in general prefer a 
positive sign for the real part of $a_{\eta' p}$.

The mass shift, Eq.(3), is very similar 
to the expectations of the Quark Meson Coupling model
(QMC, for a review see \cite{qmc}).
QCD inspired models of the $\eta$ and $\eta'$ nucleus 
systems are constructed with different selections of 
``good physics input'':
how they treat confinement, chiral symmetry and axial U(1) dynamics.
In the QMC model medium modifications are calculated at
the quark level through coupling of the light quarks in 
the hadron to the scalar isoscalar $\sigma$ 
(and also $\omega$ and $\rho$) mean fields in the nucleus.
In these calculations the large $\eta$ and $\eta'$ 
masses are used to motivate taking a MIT Bag 
description for the meson wavefunctions.
Gluonic topological effects are understood to be 
``frozen in'', meaning that they are only present 
implicitly through the masses and mixing angle in the model.
The strange-quark component of the wavefunction does not couple to the $\sigma$ field and $\eta$-$\eta'$ mixing is readily built into the model.
Possible binding energies and the in-medium masses of 
the $\eta$ and $\eta'$ are sensitive to the flavour-singlet component in the mesons and hence to the non-perturbative 
glue associated with axial U(1) dynamics \cite{etaA}.

With an $\eta$-$\eta'$ mixing angle of $-20^\circ$ 
the QMC prediction for the $\eta'$ mass in medium at nuclear matter density is 921 MeV, that is a mass shift of $-37$ MeV. 
This value is in excellent agreement with the mass shift 
$-37 \pm 10 \pm 10$ MeV deduced from photoproduction data 
\cite{nanova}.
Mixing increases the octet relative to singlet component in
the $\eta'$, reducing the binding through increased strange
quark component in the $\eta'$ wavefunction.
Without the gluonic mass contribution the $\eta'$ 
would be a strange quark state after $\eta$-$\eta'$ mixing.
Within the QMC model there would be no coupling to the 
$\sigma$ mean field and no mass shift so that any observed mass shift is induced by glue associated with the QCD axial anomaly that generates part of the $\eta'$ mass.

Increasing the flavour-singlet component in the $\eta$ at 
the expense of the octet component gives more attraction, 
more binding and a larger value of the $\eta$-nucleon scattering length, $a_{\eta N}$. 
$\eta$-$\eta'$ mixing with the phenomenological mixing angle 
$-20^\circ$ leads to a factor of two increase 
in the mass-shift and 
in the scattering length obtained in the model
relative to the prediction for a pure octet $\eta_8$.
This result may explain why values of $a_{\eta N}$ 
extracted from phenomenological fits to experimental 
data where the $\eta$-$\eta'$ mixing angle is unconstrained 
\cite{wycech} 
give larger values 
(with real part about 0.9 fm)
than those predicted 
in theoretical coupled channels models where the $\eta$ 
is treated as a pure octet state \cite{etaweise,etaoset}.

The QMC model also predicts an effective proton mass about 
755 MeV at nuclear matter density \cite{qmc} and for the
$S_{11}$ an excitation energy of $\sim 1544$ MeV \cite{etaA}, 
consistent with observations.
For the $\eta'$ in medium, larger mass shifts, downwards 
by up to 80-150 MeV, were found in recent 
Nambu-Jona-Lasinio model calculations (without confinement) 
\cite{hirenzaki} and in linear sigma model calculations 
(in a hadronic basis) \cite{jido1} which also suggest a 
rising $\eta$ effective mass at finite density.

New experiments are looking 
for possible $\eta'$ bound states in carbon using 
the (p, d) reaction at GSI \cite{gsi} 
and in photoproduction at ELSA \cite{elsap} and 
LEPS2 at SPring-8 \cite{muramatsu}.
The small $\eta'$ width in nuclei $20 \pm 5.0$ MeV at 
nuclear matter density in Eq.(3) was extracted 
from measurements of the transparency ratio for $\eta'$ photoproduction from nuclear targets \cite{elsa}
and suggests the possibility of relatively narrow bound 
$\eta'$-nucleus states accessible to experiments.
For clean observation of a bound state one needs the real part 
of the optical potential to be much bigger than the imaginary part.

COSY experiments are focussed on possible $\eta$ bound states 
in $^3$He and $^4$He \cite{wasaeta,smyrski,pawelc}.
The search for a signature of a bound state in the excitation functions for the reactions 
dd $\rightarrow ^3$He p$\pi^-$  and 
dd $\rightarrow ^3$He n$\pi^0$ 
below the threshold for the reaction  
dd $\rightarrow ^4$He $\eta$ gave a negative result and 
no narrow structure which could correspond to the $^4$He-$\eta$ mesic nucleus was found thus far \cite{skurzok}.
However, the new high statistics data collected by the 
WASA-at-COSY collaboration for the pd reaction in 2014 
gives a hope to observe a sharper state for the 
$^3$He-$\eta$ system. This is because the $^3$He-$\eta$ interaction is much stronger than the $^4$He-$\eta$ 
interaction,
which may be inferred from the much steeper rise of the total cross section at the threshold for the $\eta$ meson production via the
pd $\rightarrow ^3$He $\eta$ reaction than via 
dd $\rightarrow ^4$He $\eta$. 
It is expected that in the pessimistic case the new data will permit us to lower the upper bound for the cross section of 
the production of the $^3$He $\eta$,
{\it e.g.} via the 
pd $\rightarrow$ ($^3$He $\eta$)$_{\rm bound} \rightarrow$ 
ppp $\pi^-$  reaction from the present limit of 270 nb 
\cite{smyrski}
by about an order of magnitude. 
Such a sensitivity should 
permit us to reach the range of values of the cross section 
expected for the creation of the $\eta$-mesic $^3$He 
\cite{wilkin}.

\section{$\eta'$--$\pi$ interactions and $1^{-+}$ exotics}

Following the discussion in Section 2, the leading 
contribution to the decay 
$\eta' \rightarrow \eta \pi \pi$
within 
the QCD effective Lagrangian approach 
is associated with the OZI violating interaction
$\lambda Q^2 \partial_{\mu} \pi_a \partial^{\mu} \pi_a$
\cite{veccb}.
When iterated in the Bethe-Salpeter equation for $\eta' \pi$
rescattering this interaction yields a dynamically generated 
resonance with quantum numbers $J^{PC} = 1^{-+}$ 
and mass about 1400 MeV.
The generation of this state is mediated 
by the OZI violating coupling of the $\eta'$ \cite{bassmarco}.
One finds a possible dynamical interpretation of the 
light-mass $1^{-+}$ exotics observed in experiments at BNL and CERN \cite{exoticrev}.
This OZI violating interaction will also contribute to 
higher $L$ odd partial waves with quantum numbers $L^{-+}$.
These states are particularly interesting because 
the quantum numbers $1^{-+}, 3^{-+}, 5^{-+}$...
are inconsistent with a simple quark-antiquark bound state.
The COMPASS experiment at CERN has recently measured 
exclusive production of $\eta' \pi^-$ and $\eta \pi^-$ 
in 191 GeV $\pi^-$ collisions on a hydrogen target
\cite{compassexotic}.
They find the interesting result that $\eta' \pi^-$ 
production is enhanced relative to $\eta \pi^-$ production 
by a factor of 5-10 in the exotic $L=1,3,5$ partial waves 
with quantum numbers $L^{-+}$ in the inspected invariant 
mass range up to 3 GeV. No enhancement was observed in the
even $L$ partial waves.
We note also recent calculations where the observed light 
1$^{-+}$ states have been interpreted within the 
Dyson-Schwinger-Bethe-Salpeter framework in a quark-gluon basis \cite{krassnigg}.

\section{Conclusions}

Dedicated studies of the $\eta'$ and its interactions 
with nucleons, nuclei and other mesons have revealed 
a rich phenomenology characterised by OZI violation. 
Gluonic degrees of freedom play a vital role in
generating the $\eta'$ mass, 
medium modification of the $\eta'$ properties 
including the effective mass shift and in medium
scattering length, as well as driving decay processes
involving the $\eta'$ and dynamical generation of exotic
quantum numbers in the $\eta' \pi$ system.
Experiments using COSY-11 and at ELSA, GSI and JLab 
have taught us much about $\eta'$ production dynamics
from nucleons and nuclei and comparison of $\eta'$ interactions with the corresponding $\pi^0$ and $\eta$ processes.

\vspace{0.5cm}

{\bf Acknowledgements} \\

We thank V. Metag, M. Nanova and K. Suzuki 
for helpful discussions.
This work was supported by the 
Polish National Science Centre 
through Grant No. 2011/03/B/ST2/01847.

\vspace{0.5cm}



\begin{thebibliography}{99}
%
\bibitem{bt2013}
S. D. Bass and A. W. Thomas, 
Acta Phys. Pol. B {\bf 45} (2014) 627.
%
\bibitem{zakopane}
S. D. Bass,
Acta Phys. Pol. B {\bf 45} (2014) 2455.
%
\bibitem{shore}
G. M. Shore, {\tt hep-ph/9812354}, {\tt hep-ph/0701171}.
%
\bibitem{pawelcosy11}
P. Moskal, {\tt hep-ph/0408162}.
%
\bibitem{elsaproton}
V. Crede et al.
(CBELSA/TAPS Collaboration), 
Phys. Rev. {\bf C80} (2009) 055202; 
I. Jaegle et al. 
(CBELSA/TAPS Collaboration), 
Eur. Phys. J {\bf A47} (2011) 11; 
B. Krusche,
J. Phys. Conf. Ser. {\bf 349} (2012) 012003.
%
\bibitem{jlabproton}
M. Williams et al. 
(CLAS Collaboration), Phys. Rev. {\bf C80} (2009) 045213.
%
\bibitem{metagexa}
V. Metag, 
Hyperfine Interact. {\bf 234} (2015) 25.
%
\bibitem{compassexotic}
C. Adolph et al. (COMPASS Collaboration), 
Phys. Lett. {\bf B740} (2015) 303.
%
\bibitem{pawelrev}
P. Moskal, M. Wolke, A. Khoukaz and W. Oelert,
Prog. Part. Nucl. Phys. {\bf 49} (2002) 1.
%
\bibitem{eryk2014}
E. Czerwinski et al. (COSY-11 Collaboration),
Phys. Rev. Lett. {\bf 113} (2014) 062004.
%
\bibitem{erykwidth}
E. Czerwinski et al. 
(COSY-11 Collaboration)
Phys. Rev. Lett. {\bf 105} (2010) 122001. 
%
\bibitem{nanova}
M. Nanova et al. (CBELSA/TAPS Collaboration), 
Phys. Lett. {\bf B727} (2013) 417. 
%
\bibitem{etaA}
S. D. Bass and A. W. Thomas, Phys. Lett. {\bf B634} (2006) 368.
%
\bibitem{exa14}
S. D. Bass,
Hyperfine Interact. {\bf 234} (2015) 41.
%
\bibitem{wycech}
A. M. Green and S. Wycech,
Phys. Rev. {\bf C71} (2005) 014001;
R. A. Arndt et al., Phys. Rev. {\bf C72} (2005) 045202.
%
\bibitem{elsa}
M. Nanova et al. 
(CBELSA/TAPS Collaboration), 
Phys. Lett. {\bf B710} (2012) 600.
%
\bibitem{wasaeta}
P. Adlarson et al. (WASA-at-COSY Collaboration), 
Phys. Rev. {\bf C87} (2013) 035204.
%
\bibitem{bassmarco}
S.D. Bass and E. Marco, Phys. Rev. {\bf D65} (2002) 057503.
%
\bibitem{crewther}
R. J. Crewther, Acta Physica Austriaca Suppl. {\bf 19} (1978) 47.
%
\bibitem{ks}
H. Fritzsch and P. Minkowski, Nuovo Cim. {\bf A30} (1975) 393;
J. Kogut and L. Susskind, Phys. Rev. {\bf D11} (1975) 3594; 
E. Witten, Nucl. Phys. {\bf B149} (1979) 285; 
I. Horvath, N. Isgur, J. McCune and H. B. Thacker,
Phys. Rev. {\bf D65} (2001) 014502;
R. Alkofer, C. S. Fischer and R. Williams, 
Eur. Phys. J. {\bf A38} (2008) 53.
%
\bibitem{thooft}
G. 't Hooft, Phys. Rev. Lett. {\bf 37} (1976) 8; 
Phys. Rev. {\bf D14} (1976) 3432.
%
\bibitem{witten}
E. Witten, Nucl. Phys. {\bf B156} (1979) 269.
%
\bibitem{venez}
G. Veneziano, Nucl. Phys. {\bf B159} (1979) 213.
%
\bibitem{gilman}
F. J. Gilman and R. Kauffman, Phys. Rev. {\bf D36} (1987) 2761;
(E) {\bf D37} (1988) 3348;
P. Ball, J. M. Frere and M. Tytgat, Phys. Lett. {\bf B365} (1996) 367;
T. Feldmann and P. Kroll, Phys. Scripta T {\bf 99} (2002) 13; 
F. Ambrosino et al. (KLOE Collaboration), JHEP {\bf 0907} (2009) 105.
%
\bibitem{veccb}
P. Di Vecchia, F. Nicodemi, R. Pettorino and G. Veneziano, 
Nucl. Phys. {\bf B181} (1981) 318.
%
\bibitem{hf}
H. Fritzsch, Phys. Lett. {\bf B415} (1997) 83.
%
\bibitem{bass99}
S. D. Bass,
Phys. Lett. {\bf B463} (1999) 286; {\tt hep-ph/0006348}.
%
\bibitem{tgv}
G. M. Shore and G. Veneziano, Nucl. Phys. {\bf B381} (1992) 23. 
%
\bibitem{SBrmp1}
S. D. Bass, Rev. Mod. Phys. {\bf 77} (2005) 1257.
%
\bibitem{SBrmp2}
C. A. Aidala, S. D. Bass, D. Hasch and G. K. Mallot,
Rev. Mod. Phys. {\bf 85} (2013) 655.
%
\bibitem{sbcracow}
S. D. Bass, Acta Phys. Polon. Supp. {\bf 2} (2009) 11, 
{\tt arXiv:0812.5047 [hep-ph]}.
%
\bibitem{vecca}
P. Di Vecchia and G. Veneziano, Nucl. Phys. {\bf B171}
(1980) 253;
C. Rosenzweig, J. Schechter and G. Trahern, 
Phys. Rev. {\bf D21} (1980) 3388;
E. Witten, Ann. of Phys. {\bf 128} (1981) 363;
P. Nath and R. Arnowitt, Phys. Rev. {\bf D23} (1981) 473.
%
\bibitem{paweletapa}
%
P. Moskal et al. 
(COSY-11 Collaboration), Phys. Rev. Lett. {\bf 80} (1998) 3202
%
\bibitem{paweletapb}
P. Moskal et al. 
(COSY-11 Collaboration), Phys. Lett. {\bf B474} (2000) 416
%
\bibitem{paweletapc}
P. Moskal et al. 
(COSY-11 Collaboration), Phys. Lett. {\bf B482} (2000) 356
%
\bibitem{paweletapd}
A. Khoukaz et al. 
(COSY-11 Collaboration), Eur. Phys. J. {\bf A20} (2004) 345
%
\bibitem{pklaja}
E. Czerwinski, P. Moskal and M. Silarski,
Acta Phys. Polon. B {\bf 45} (2014) 739; 
P. Klaja et al.
(COSY-11 Collaboration), Phys. Lett. {\bf B684} (2010) 11.
%
\bibitem{hibou}
F. Hibou et al.
(SPESIII Collaboration), Phys. Lett. {\bf B438} (1998) 41.
%
\bibitem{disto}
F. Balestra et al. (DISTO Collaboration),
Phys. Lett. {\bf B491} (2000) 29
%
\bibitem{faeldt}
G. F\"aldt, T. Johansson and C. Wilkin, 
Phys. Scripta T {\bf 99} (2002) 146. 
%
\bibitem{kanzo}
K. Nakayama, H. F. Arellano, J. W. Durso and J. Speth,
Phys. Rev. {\bf C61} (2000) 024001; 
F. Huang, H. Haberzettl and K. Nakayama,
Phys. Rev. {\bf C87} (2013) 054004.
%
\bibitem{heetap}
D. Barberis et al.
(WA102 Collaboration), Phys. Lett. {\bf B427} (1998) 398.
%
\bibitem{pawela}
P. Moskal et al. (COSY-11 Collaboration),
Phys. Rev. {\bf C79} (2009) 015208.
%
\bibitem{pawelb}
J. Klaja et al. (COSY-11 Collaboration), 
Phys. Rev. {\bf C81} (2010) 035209.
%
\bibitem{uppsala}
H. Calen et al. 
(CELSIUS Collaboration), Phys. Rev. {\bf C58} (1998) 2667.
%
\bibitem{paweleta}
R. Czyzykiewicz et al.
(COSY-11 Collaboration), Phys. Rev. Lett. {\bf 98} (2007) 122003. 
%
\bibitem{qfree}
P. Moskal et al.
(COSY-11 Collaboration), J. Phys. {\bf G32} (2006) 629.
%
\bibitem{kampfer}
L. P. Kaptari and B. K\"ampfer,
Eur. Phys. J. {\bf A37} (2008) 69.
%
\bibitem{kienle}
P. Kienle and T. Yamazaki, Prog. Part. Nucl. Phys. {\bf 52} (2004) 85.
%
\bibitem{suzuki}
K. Suzuki et al., Phys. Rev. Lett. {\bf 92} (2004) 072302.
%
\bibitem{smyrski}
P. Moskal and J. Smyrski, Acta Phys. Pol. B {\bf 41} (2010) 2281.
%
\bibitem{pawelc}
A. Budzanowski et al. (COSY-GEM Collaboration), 
Phys. Rev. {\bf C79} (2009) 012201;
A. Khoukaz, 
Acta Phys. Polon. B {\bf 45} (2014) 655;
W. Krzemien, P. Moskal and M. Skurzok, 
Acta Phys. Polon. B {\bf 45} (2014) 689;
W. Krzemien, P. Moskal and M. Skurzok,
Acta Phys. Polon. B {\bf 46} (2015) 757.
%
\bibitem{krusche}
B. Krusche and C. Wilkin, Prog. Part. Nucl. Phys. {\bf 80} (2014) 43;
H. Machner, J. Phys. {\bf G42} (2015) 043001.
%
\bibitem{gsi}
K. Itahashi, H. Fujioka et al., Prog. Theor. Phys. {\bf 128} (2012) 601.
%
\bibitem{elsap}
V. Metag et al., approved ELSA/03-2012-BGO-OD.
%
\bibitem{muramatsu}
N. Muramtsu, {\tt arXiv:1307.6411 [physics.ins-det]}.
%
\bibitem{mamieta}
M. Pfeiffer et al., Phys. Rev. Lett. {\bf 92} (2004) 252001;
F. Pheron et al., Phys. Lett. {\bf B709} (2012) 21.
%
\bibitem{pdeta}
%
J. Smyrski et al.
(COSY-11 Collaboration), Phys. Lett. {\bf B649} (2007) 258;
T. Mersmann et al.
(ANKE Collaboration), Phys. Rev. Lett. {\bf 98} (2007) 242301.
%
\bibitem{etaqmc}
K. Tsushima, D. H. Lu, A. W. Thomas and K. Saito,
Phys. Lett. {\bf B443} (1998) 26; 
K. Tsushima, Nucl. Phys. {\bf A670} (2000) 198c.
%
\bibitem{wilkin}
C. Wilkin, Acta Phys. Pol. B {\bf 45} (2014) 603.
%
\bibitem{etaArevs}
Q. Haider and L. C. Liu, {\tt arXiv:1509.05487 [nucl-th]};
N. G. Kelkar, K.P. Khemchandani, N.J. Upadhyay and B.K. Jain,
Rept. Prog. Phys. {\bf 76} (2013) 066301;
N. G. Kelkar, Acta Phys. Polon. B {\bf 46} (2015) 113.
%
\bibitem{hirenzaki} 
H. Nagahiro, M. Takizawa and S. Hirenzaki, 
Phys. Rev. {\bf C74} (2006) 045203.
%
\bibitem{jido1} 
D. Jido, H. Nagahiro and S. Hirenzaki, 
Phys. Rev. {\bf C85} (2012) 032201;
%
H. Nagahiro et al., Phys. Rev. {\bf C87} (2013) 045201.
%
\bibitem{gal1}
E. Friedman, A. Gal and J. Mares,
Phys. Lett. {\bf B725} (2013) 334; 
A. Cieply, E. Friedman, A. Gal and J. Mares,  
Nucl. Phys. {\bf A925} (2014) 126;
N. Barnea, E. Friedman and A. Gal, Phys. Lett. {\bf B747} (2015) 345.
%
\bibitem{gillitzer}
A. Schr\"oter et al., Z Physik {\bf A350} (1994) 101.
%
\bibitem{kaons}
R. Barth et al. (KaoS Collaboration), 
Phys. Rev. Lett. {\bf 78} (1997) 4007.
%
\bibitem{averbeck}
R. Averbeck et al., Z Physik {\bf A359} (1997) 65.
%
\bibitem{robig}
M. R\"obig-Landau et al., Phys. Lett. {\bf B373} (1996) 45.
%
\bibitem{yorita}
T. Yorita et al., Phys. Lett. {\bf B476} (2000) 226.
%
\bibitem{metag}
J. Weil, U. Mosel and V. Metag, Phys. Lett. {\bf B723} (2013) 120; 
V. Metag et al., Prog. Part. Nucl. Phys. {\bf 67} (2012) 530.
%
\bibitem{ericson}
T. E. O. Ericson and W. Weise, {\it Pions and Nuclei},
Oxford UP (1988).
%
\bibitem{qmc}
K. Saito, K. Tsushima and A. W. Thomas, 
Prog. Part. Nucl. Phys. {\bf 58} (2007) 1;
%
P.~A.~M.~Guichon, K.~Saito, E.~N.~Rodionov and A.~W.~Thomas,
  Nucl. Phys. {\bf A601} (1996) 349;
  P.~A.~M.~Guichon, Phys. Lett. {\bf B200} (1988) 235.
%
\bibitem{etaweise}
N. Kaiser, T. Waas and W. Weise, Nucl. Phys. {\bf A612} 
(1997) 297;
T. Waas and W. Weise, Nucl. Phys. {\bf A625} (1997) 287.
%
\bibitem{etaoset}
T. Inoue and E. Oset, Nucl. Phys. {\bf A710} (2002) 354;
C. Garcia-Recio, T. Inoue, J. Nieves and E. Oset,
Phys. Lett. {\bf B550} (2002) 47.
%
\bibitem{skurzok}
M. Skurzok et al., these proceedings. 
%
\bibitem{exoticrev}
C. A. Meyer and E. S. Swanson,
Prog. Part. Nucl. Phys. {\bf 82} (2015) 21.
%
\bibitem{krassnigg}
T. Hilger, M. Gomez-Rocha and A. Krassnigg,
Phys. Rev. {\bf D91} (2015) 114004.
%
\end{thebibliography}
\end{document}